\documentclass[preprint,showpacs,superscriptaddress,amsmath,amssymb]{revtex4}

\usepackage{graphicx}
\usepackage{dcolumn}
\usepackage{bm}

\def\ra     {\rightarrow}
\def\mm     {\mu^{+}\mu^{-}}
\def\jpsi   {J/\psi}
\def\jpsimm {\jpsi\ra\mm}
\def\Bp     {B^{\pm}}
\def\jpsik  {\jpsi K^{\pm}}
\def\jpsip  {\jpsi \pi^{\pm}}
\def\Bpjk   {\Bp\ra\jpsik}
\def\Bpjp   {\Bp\ra\jpsip}
\def\ratio  {\mathcal{B}(\Bpjp)/\mathcal{B}(\Bpjk)}

\begin{document}

\preprint{FERMILAB-PUB-07-170-E}

\title{\bf\boldmath Measurement of the Ratio of Branching Fractions $\ratio$}

\affiliation{Institute of Physics, Academia Sinica, Taipei, Taiwan 11529, Republic of China} 
\affiliation{Argonne National Laboratory, Argonne, Illinois 60439} 
\affiliation{Institut de Fisica d'Altes Energies, Universitat Autonoma de Barcelona, E-08193, Bellaterra (Barcelona), Spain} 
\affiliation{Baylor University, Waco, Texas  76798} 
\affiliation{Istituto Nazionale di Fisica Nucleare, University of Bologna, I-40127 Bologna, Italy} 
\affiliation{Brandeis University, Waltham, Massachusetts 02254} 
\affiliation{University of California, Davis, Davis, California  95616} 
\affiliation{University of California, Los Angeles, Los Angeles, California  90024} 
\affiliation{University of California, San Diego, La Jolla, California  92093} 
\affiliation{University of California, Santa Barbara, Santa Barbara, California 93106} 
\affiliation{Instituto de Fisica de Cantabria, CSIC-University of Cantabria, 39005 Santander, Spain} 
\affiliation{Carnegie Mellon University, Pittsburgh, PA  15213} 
\affiliation{Enrico Fermi Institute, University of Chicago, Chicago, Illinois 60637} 
\affiliation{Comenius University, 842 48 Bratislava, Slovakia; Institute of Experimental Physics, 040 01 Kosice, Slovakia} 
\affiliation{Joint Institute for Nuclear Research, RU-141980 Dubna, Russia} 
\affiliation{Duke University, Durham, North Carolina  27708} 
\affiliation{Fermi National Accelerator Laboratory, Batavia, Illinois 60510} 
\affiliation{University of Florida, Gainesville, Florida  32611} 
\affiliation{Laboratori Nazionali di Frascati, Istituto Nazionale di Fisica Nucleare, I-00044 Frascati, Italy} 
\affiliation{University of Geneva, CH-1211 Geneva 4, Switzerland} 
\affiliation{Glasgow University, Glasgow G12 8QQ, United Kingdom} 
\affiliation{Harvard University, Cambridge, Massachusetts 02138} 
\affiliation{Division of High Energy Physics, Department of Physics, University of Helsinki and Helsinki Institute of Physics, FIN-00014, Helsinki, Finland} 
\affiliation{University of Illinois, Urbana, Illinois 61801} 
\affiliation{The Johns Hopkins University, Baltimore, Maryland 21218} 
\affiliation{Institut f\"{u}r Experimentelle Kernphysik, Universit\"{a}t Karlsruhe, 76128 Karlsruhe, Germany} 
\affiliation{High Energy Accelerator Research Organization (KEK), Tsukuba, Ibaraki 305, Japan} 
\affiliation{Center for High Energy Physics: Kyungpook National University, Taegu 702-701, Korea; Seoul National University, Seoul 151-742, Korea; and SungKyunKwan University, Suwon 440-746, Korea} 
\affiliation{Ernest Orlando Lawrence Berkeley National Laboratory, Berkeley, California 94720} 
\affiliation{University of Liverpool, Liverpool L69 7ZE, United Kingdom} 
\affiliation{University College London, London WC1E 6BT, United Kingdom} 
\affiliation{Centro de Investigaciones Energeticas Medioambientales y Tecnologicas, E-28040 Madrid, Spain} 
\affiliation{Massachusetts Institute of Technology, Cambridge, Massachusetts  02139} 
\affiliation{Institute of Particle Physics: McGill University, Montr\'{e}al, Canada H3A~2T8; and University of Toronto, Toronto, Canada M5S~1A7} 
\affiliation{University of Michigan, Ann Arbor, Michigan 48109} 
\affiliation{Michigan State University, East Lansing, Michigan  48824} 
\affiliation{Institution for Theoretical and Experimental Physics, ITEP, Moscow 117259, Russia} 
\affiliation{University of New Mexico, Albuquerque, New Mexico 87131} 
\affiliation{Northwestern University, Evanston, Illinois  60208} 
\affiliation{The Ohio State University, Columbus, Ohio  43210} 
\affiliation{Okayama University, Okayama 700-8530, Japan} 
\affiliation{Osaka City University, Osaka 588, Japan} 
\affiliation{University of Oxford, Oxford OX1 3RH, United Kingdom} 
\affiliation{University of Padova, Istituto Nazionale di Fisica Nucleare, Sezione di Padova-Trento, I-35131 Padova, Italy} 
\affiliation{LPNHE, Universite Pierre et Marie Curie/IN2P3-CNRS, UMR7585, Paris, F-75252 France} 
\affiliation{University of Pennsylvania, Philadelphia, Pennsylvania 19104} 
\affiliation{Istituto Nazionale di Fisica Nucleare Pisa, Universities of Pisa, Siena and Scuola Normale Superiore, I-56127 Pisa, Italy} 
\affiliation{University of Pittsburgh, Pittsburgh, Pennsylvania 15260} 
\affiliation{Purdue University, West Lafayette, Indiana 47907} 
\affiliation{University of Rochester, Rochester, New York 14627} 
\affiliation{The Rockefeller University, New York, New York 10021} 
\affiliation{Istituto Nazionale di Fisica Nucleare, Sezione di Roma 1, University of Rome ``La Sapienza," I-00185 Roma, Italy} 
\affiliation{Rutgers University, Piscataway, New Jersey 08855} 
\affiliation{Texas A\&M University, College Station, Texas 77843} 
\affiliation{Istituto Nazionale di Fisica Nucleare, University of Trieste/\ Udine, Italy} 
\affiliation{University of Tsukuba, Tsukuba, Ibaraki 305, Japan} 
\affiliation{Tufts University, Medford, Massachusetts 02155} 
\affiliation{Waseda University, Tokyo 169, Japan} 
\affiliation{Wayne State University, Detroit, Michigan  48201} 
\affiliation{University of Wisconsin, Madison, Wisconsin 53706} 
\affiliation{Yale University, New Haven, Connecticut 06520} 
\author{A.~Abulencia}
\affiliation{University of Illinois, Urbana, Illinois 61801}
\author{J.~Adelman}
\affiliation{Enrico Fermi Institute, University of Chicago, Chicago, Illinois 60637}
\author{T.~Affolder}
\affiliation{University of California, Santa Barbara, Santa Barbara, California 93106}
\author{T.~Akimoto}
\affiliation{University of Tsukuba, Tsukuba, Ibaraki 305, Japan}
\author{M.G.~Albrow}
\affiliation{Fermi National Accelerator Laboratory, Batavia, Illinois 60510}
\author{D.~Ambrose}
\affiliation{Fermi National Accelerator Laboratory, Batavia, Illinois 60510}
\author{S.~Amerio}
\affiliation{University of Padova, Istituto Nazionale di Fisica Nucleare, Sezione di Padova-Trento, I-35131 Padova, Italy}
\author{D.~Amidei}
\affiliation{University of Michigan, Ann Arbor, Michigan 48109}
\author{A.~Anastassov}
\affiliation{Rutgers University, Piscataway, New Jersey 08855}
\author{K.~Anikeev}
\affiliation{Fermi National Accelerator Laboratory, Batavia, Illinois 60510}
% ...
\collaboration{CDF Collaboration\footnote{With visitors from $^a$University of Athens, 
$^b$University of Bristol, 
$^c$University Libre de Bruxelles, 
$^d$Cornell University, 
$^e$University of Cyprus, 
$^f$University of Dublin, 
$^g$University of Edinburgh, 
$^h$University of Heidelberg, 
$^i$Universidad Iberoamericana, 
$^j$University of Manchester, 
$^k$Nagasaki Institute of Applied Science, 
$^l$University de Oviedo, 
$^m$University of London, Queen Mary and Westfield College, 
$^n$Texas Tech University, 
$^o$IFIC(CSIC-Universitat de Valencia), 
}}
\noaffiliation

\date{\today}

\begin{abstract}
We report a measurement of the ratio of branching fractions of the decays $\Bpjp$ and $\Bpjk$ using the 
CDF II detector at the Fermilab Tevatron Collider. 
The signal from the Cabbibo-suppressed $\Bpjp$ decay is separated from $\Bpjk$ using the $\Bpjk$ 
invariant mass distribution and the kinematical differences of the hadron track in the two decay modes. 
From a sample of 220 pb$^{-1}$ of $p\bar{p}$ collisions at $\sqrt{s}=\mbox{1.96 TeV}$, we observe 
91$\pm$15 $\Bpjp$ events together with 1883$\pm$34 $\Bpjk$ events. 
The ratio of branching fractions is found to be 
$\mathcal{B}(\Bpjp)/\mathcal{B}(\Bpjk) = (4.86\pm0.82(\mbox{stat.})\pm0.15(\mbox{syst.}))\%$.
\end{abstract}


\maketitle

\newpage
The $\Bpjp$ decay is a Cabibbo-suppressed mode proceeding via a $b \ra c\bar{c}d$ transition. 
If the leading-order tree diagram is the dominant contribution, its branching fraction is expected to 
be $\approx$ 5\% of that of the Cabibbo-favored mode $\Bpjk$. 
Detailed predictions of the ratio are obtained using the hypothesis of factorization of the hadronic 
matrix elements~\cite{wirbel,bauer}, a theoretical approach widely used in the treatment of 
non-leptonic decays of $B$ mesons. 
However, the absence of strong theoretical arguments supporting factorization and the use of 
phenomenological models, which are a source of theoretical uncertainties, weakens the reliability of 
those predictions, which need to be accurately tested on data. 
Until now, the measurements on the $\Bpjp$ decay were performed by many experiments. 
The BABAR collaboration reported $\ratio=(5.37\pm0.45)\%$ with $244\pm20$ $\Bpjp$ events~\cite{babar2}. 
The Belle collaboration reported $\mathcal{B}(\Bpjp)=(3.8 \pm 0.6)\times 10^{-5}$~\cite{belle}. 
A previous study of the $\Bpjp$ decay was also performed by the CLEO collaboration~\cite{cleo}. 
The result of this analysis supersedes the previous CDF result~\cite{cdf}. 

This paper presents a measurement of the ratio of branching fractions $\ratio$. 
We use a sample of fully reconstructed $\Bpjk$ decays, where $\jpsimm$, corresponding to an integrated 
luminosity of 220 pb$^{-1}$ of $p\bar{p}$ collisions at $\sqrt{s}=$1.96 TeV collected by the CDF II 
detector at Fermilab between February 2002 and August 2003. 

The CDF II detector is a multipurpose detector~\cite{acosta} with a central geometry and has a tracking 
system surrounded by calorimeters and muon detectors. 
The components of the detector most relevant to this analysis are described briefly here. 
Charged particle trajectories are reconstructed in the pseudorapidity range $|\eta|<\mbox{1.0}$, where 
$\eta = - \ln (\tan\frac{\theta}{2})$ and $\theta$ is the polar angle measured from the beam 
line~\cite{geom}. 
Trajectories are reconstructed from hits in the silicon microstrip detector (SVX II)~\cite{sill} and 
the central outer tracker (COT)~\cite{affolder} which are immersed in a 1.4 T solenoidal magnetic field. 
The SVX II consists of five concentric layers made of double-sided silicon detectors with radii between 
2.5 and 10.6 cm, each providing a position measurement with 15 $\mu$m resolution in the $r$-$\phi$ plane. 
The COT is an open-cell drift chamber with 96 measurement layers, between 40 and 137 cm in radius, 
organized into eight alternating axial and $\pm\mbox{2}^{\circ}$ stereo superlayers. 
The transverse momentum ($p_{T}$) resolution is 
$\sigma_{p_{T}}/p_{T}\simeq 0.15\%p_{T}(\mbox{GeV/}c)^{-1}$. 
Muon detectors consisting of multi-layer drift chambers are located radially around the outside of the 
calorimeter~\cite{ascoli}. 
The central muon detector (CMU) covers a range in pseudorapidity of $|\eta| <$ 0.6. 
The central muon extension (CMX) extends the pseudorapidity coverage to $\mbox{0.6}<|\eta|<\mbox{1.0}$. 

The data sample used in this analysis required a dimuon trigger sensitive to $\jpsimm$. 
The CDF II detector employs a three-level trigger system to select events of interest efficiently. 
At the first trigger level, muon candidates are identified by matching track segments in the CMU and CMX 
to coarsely reconstructed COT tracks obtained with the extremely fast tracker (XFT)~\cite{thomson}. 
Dimuon triggers use combinations of CMU-CMU and CMU-CMX muons with $p_{T}>$1.5 (2.0) $\mbox{GeV/}c$ for 
CMU (CMX) muons. 
For the data presented here, no additional requirements are made at the second level. 
At the third trigger level, a detailed reconstruction is performed and oppositely charged dimuon events 
with an invariant mass in the range of $\mbox{2.7}-\mbox{4.0 GeV}/c^{2}$ are selected. 

In this analysis, we reconstruct $\Bpjk$ decays. 
$B$ meson decay modes involving the well-known $\jpsimm$ decay have been extensively used in other 
measurements at CDF, and their selection criteria are well established. 
We follow the selection requirements developed in the $b$ hadron mass measurement~\cite{korn} and apply 
them to the $\Bp$ decay mode of interest. 
To ensure the best momentum scale calibration, the data sample used for this analysis is also kept the 
same as that for the mass measurement.  

The $\Bpjk$ reconstruction begins by selecting $\jpsimm$ candidates with pairs of oppositely charged 
tracks which satisfy the requirements of the dimuon triggers. 
$\jpsi$ candidates are further selected by requiring their invariant mass to be within 
$\mbox{80 MeV}/c^{2}$ of the world average $\jpsi$ mass~\cite{pdg}. 
After a $\jpsi$ candidate is identified, any other charged track is assumed to be a kaon and is 
combined with the $\jpsi$ candidate to make a $\Bp$ candidate. 
The tracks of the kaon and two muons are then fitted to a common three dimensional vertex (3-D) while 
constraining the invariant mass of two muons to the world average $\jpsi$ mass~\cite{pdg}. 
To ensure good vertex resolution, each track must have hits in at least three silicon vertex detector 
layers in the $r-\phi$ plane and the probability resulting from the 3-D vertex fit is required to be 
greater than 1\%. 

A number of further requirements are made to improve the signal-to-background separation. 
Prompt background, with tracks coming directly from the primary vertex, can be reduced by exploiting 
variables sensitive to the long lifetime of the $\Bp$ meson. 
To reduce prompt background, the transverse decay length ($L_{xy}$) of the $\Bp$ is required to exceed 
200 $\mu$m, where $L_{xy}$ is defined as the vector from the primary vertex to the $\Bp$ decay vertex 
projected onto the $p_{T}$ of the $\Bp$ candidate. 
To further reduce combinatorial background, we require $p_{T} > 6.5\mbox{ GeV/}c$ for the $\Bp$ 
candidate and $p_{T} > 2.0\mbox{ GeV/}c$ for the hadron from the $\Bp$ decay. 
The values used in the above selection criteria are determined by an iterative optimization procedure 
in which the significance $S/\sqrt{S+B}$ is maximized. 
The quantity $S$ represents the number of accepted signal events, in this case taken from a Monte Carlo 
simulation sample, and $B$ is the number of selected $\Bp$ candidates within the mass sidebands of 
the data. 

We measure the following ratio: 
\begin{align}
\frac{\mathcal{B}(\Bpjp)}{\mathcal{B}(\Bpjk)}&=\frac{N_{\jpsip}}{N_{\jpsik}}\times\frac{\epsilon_{\jpsik}}{\epsilon_{\jpsip}} \nonumber \\
&=r_{obs}\times\frac{1}{\epsilon_{rel}}, 
\label{eq:ratio}
\end{align}
where $r_{obs}(\equiv N_{\jpsip}/N_{\jpsik})$ is the ratio of the yields of each decay mode, and 
$\epsilon_{rel}(\equiv \epsilon_{\jpsip}/\epsilon_{\jpsik})$ is the relative reconstruction efficiency. 
In this analysis, the quantity $r_{obs}$ is extracted from an unbinned maximum likelihood fit using the 
differences between the two decay modes in the mass distribution, and is corrected with $\epsilon_{rel}$ 
obtained from Monte Carlo simulation. 

\begin{figure}[htbp]
\begin{center}
\includegraphics[width=8cm]{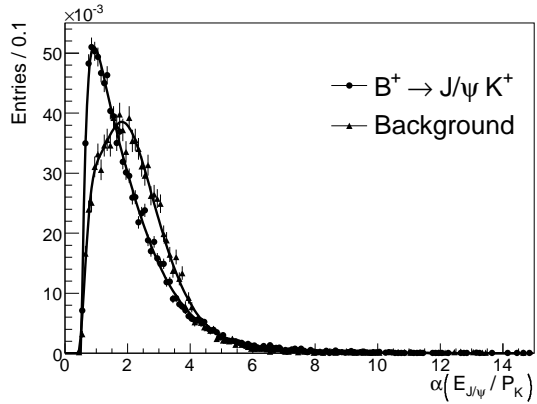}
\caption{\label{fig:alpha} The $\alpha$ distributions of $\Bpjk$, which are obtained with 
Monte Carlo simulation, and background obtained from the non-signal data sample. 
The solid curves are the corresponding parameterization functions from Eq.~(\ref{eq:alphas}) and (\ref{eq:alphab}). 
The $\alpha$ distributions of the two signals are very similar in shape due to the almost identical 
kinematics of the two decay modes.
To avoid confusion from it, we plot the $\alpha$ distribution of $\Bpjk$ only.}
\end{center}
\end{figure}

To build the probability density function (PDF) used in the unbinned maximum likelihood fit, we choose 
the invariant mass of $J/\psi$ and a kaon ($M_{J/\psi K}$) as an observable. 
There are three components in the distribution of the $M_{J/\psi K}$ variable: the $\Bpjk$ signal, the 
$\Bpjp$ signal and the combinatorial background. 
As demonstrated in the high statistics $D$ and $B$ mass reconstuctions with similar decay topology, the 
invariant mass distribution of  $\Bpjk$ decay at CDF  is  well described by 
a Gaussian function with width determined by CDF's tracking resolution~\cite{korn}~\cite{bhh}~\cite{Dmass}.  
Therefore, we model the $\Bpjk$ signal as a Gaussian centered at the mass of $\Bp$ ($M_{B}$) with a width 
$\sigma_{K}$. 
If the pion mass were assigned to the hadron track originating from the $\Bpjp$ decay, the resulting 
spectrum would be also a Gaussian centered at $M_{B}$. 
However, assigning the kaon mass to this track produces a spectrum partially overlapping the $\Bpjk$ 
and shifted in the positive direction. 
The shifted invariant mass of $\Bpjp$ can be calculated by an approximation, which has a good agreement 
with the exact value~\cite{Mass-alpha}, 
\begin{equation}
\mathcal{M}_{B}^{\prime 2}(\alpha) \simeq M_{B}^{2}+(1+\alpha)(M_{K}^{2}-M_{\pi}^{2}), 
\label{eq:camass}
\end{equation}
where $M_{K}$ and $M_{\pi}$ are respectively the kaon and the pion masses. 
The purely kinematic variable $\alpha$ is defined as $\alpha\equiv E_{\jpsi}/P_{K}$, where $E_{\jpsi}$ 
is the $\jpsi$ energy and $P_{K}$ is the magnitude of the momentum of the hadron track. 
Using Eq.~(\ref{eq:camass}), the $\Bpjp$ signal is modeled as a Gaussian centered at 
$\mathcal{M}_{B}^{\prime}(\alpha)$ with a width $\sigma_{\pi}$. 
We find $\sigma_{K}$ and $\sigma_{\pi}$ have almost the same value from the Monte Carlo simulation, so 
we constrain them to be the same value in the fit.
We assume the background mass distribution is a first order polynomial. 
In the likelihood, we also include the PDF functions of $\alpha$ for $\Bpjk$ and $\Bpjp$ as the distributions 
for the two signals are found to be slightly different.
We parameterize  $\alpha$ PDF distributions from Monte Carlo simulation. 
We also parameterize the $\alpha$ distribution of the background, which is obtained from the mass 
sidebands of the data. 
These mass sidebands are chosen from $\mbox{5.2}<M_{J/\psi K}<\mbox{5.24}$ and 
$\mbox{5.4}<M_{J/\psi K}<\mbox{5.6 GeV/}c^{2}$ to avoid signal contaminations and other backgrounds 
from partially reconstructed $B$ mesons that fall below 5.2 GeV/$c^{2}$. 
The empirical functions used in the parameterizations are 
\begin{align}
& h_{\jpsi X}(\alpha;f_{i},\lambda_{i},a)=\sum_{i=1}^{3} f_{i} (\alpha-a) e^{-\lambda_{i}\alpha} \label{eq:alphas}\\
& h_{bkg}(\alpha;f_{i},\lambda_{i},a)=\sum_{i=1}^{3} f_{i} (\alpha-a)^{3} e^{-\lambda_{i}\alpha} \label{eq:alphab}, 
\end{align}
where the symbol $X$ denotes $K$ or $\pi$ in Eq.~(\ref{eq:alphas}) and $f_{1}$, $f_{2}$ and $f_{3}$ are 
to be the fractional contributions of each type of function when the functions are properly normalized 
to 1. 
Because of the requirement on the $p_{T}$ of the hadron track and also of the dimuon triggers, all 
$\alpha$ distributions show a cutoff around 0.5 in the $\alpha$ variable and these cutoff values are 
parameterized by $a$ in Eq.~(\ref{eq:alphas}) and (\ref{eq:alphab}). 
These parameters of the functions describing the $\alpha$ distributions are fixed in the fit. 
The $\alpha$ distributions of the signal and background, and the results of the parameters are shown in 
Fig.~\ref{fig:alpha}. 
With models for each signal and background, and with the chosen observables, the PDF of the $i^{th}$ 
event is written as 
\begin{widetext}
\begin{align}
p_{i}= & f_{s} \left[\frac{1}{1+r_{obs}}G\left(M_{J/\psi K}^{i}-M_{B}, \sigma\right)h_{\jpsi K}(\alpha^{i}) +
                     \frac{r_{obs}}{1+r_{obs}}G\left(M_{J/\psi K}^{i}-\mathcal{M}_{B}^{\prime}(\alpha^{i}),\sigma\right)h_{\jpsi \pi}(\alpha^{i})
               \right] \nonumber\\
       & + (1-f_{s})B(M_{J/\psi K}^{i})h_{bkg}(\alpha^{i}),
\label{eq:like}
\end{align}
\end{widetext}
where $f_{s}$ is the fraction of signal events in the data sample and $r_{obs}$ is the ratio between 
the yields of each signal. 
The functions, $G\left(M_{J/\psi K}^{i}-M_{B}, \sigma\right)$
and $G\left(M_{J/\psi K}^{i}-\mathcal{M}_{B}^{\prime}(\alpha^{i}), \sigma\right)$, are Gaussians with 
a width $\sigma$ describing the mass distributions of $\Bpjk$ and $\Bpjp$, respectively, and 
$B(M_{J/\psi K}^{i})$ is a first order polynomial function which describes the background mass 
distribution. 
The fitting range ($\mbox{5.2}<M_{J/\psi K}<\mbox{5.6 GeV/}c^{2}$) is selected to avoid the 
backgrounds from partially reconstructed $B$ mesons, but to include enough of the background region to 
determine accurately the background shape. 
$\mathcal{L}=\prod_{i=1}^N p_i$ is then maximized to obtain the best fit values for 
$M_{B}$, $\sigma$, $f_{s}$ and $r_{obs}$.  The fitter is extensively tested with Monte Carlo samples. 

\begin{figure}[htbp]
\begin{center}
\includegraphics[width=8cm]{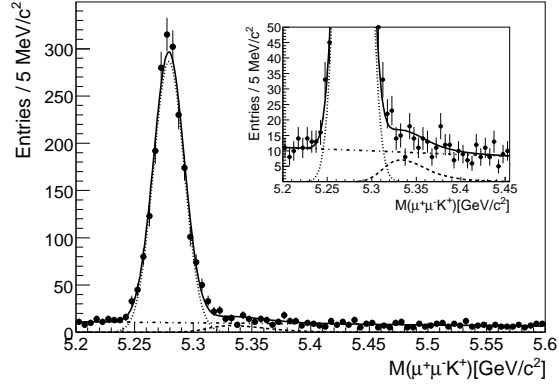}
\caption{\label{fig:mover} The invariant mass distribution in the data sample (points) projected
with the results of the likelihood fit; overall (solid line), $\Bpjk$ (dotted line), 
$\Bpjp$ (dashed line) and background (dashed-dotted line). The inset shows the magnified region of the 
$\Bpjp$ signal.}
\end{center}
\end{figure}

\begin{figure}[htbp]
\begin{center}
\includegraphics[width=8cm]{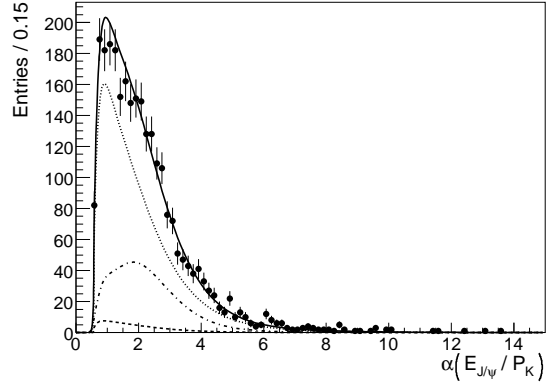}
\caption{\label{fig:aover} The $\alpha$ distribution in the data sample (points) compared 
with the results of the likelihood fit; overall (solid line), $\Bpjk$ (dotted line),
$\Bpjp$ (dashed line) and background (dashed-dotted line).}
\end{center}
\end{figure}

The fit to 2683 candidates falling in the fitting range returns the signal fraction, 
$f_{s}=\mbox{0.736}\pm\mbox{0.012}$, and the ratio of the yields of each decay mode, 
$r_{obs}=(\mbox{4.82}\pm\mbox{0.81})\%$. 
These values give 1883$\pm$34 signal events in the $\Bpjk$ decay mode and 91$\pm$15 events in the 
$\Bpjp$ decay mode. 
The distributions in $M_{J/\psi K}$ and $\alpha$ for the events in the data sample are shown in 
Fig.~\ref{fig:mover} and Fig.~\ref{fig:aover}, along with the likelihood fit results. 

Possible biases in the fitting procedure are investigated by performing the fit on Monte Carlo samples generated 
by the PDF in Eq.~(\ref{eq:like}), with known composition and with the same size as the data sample. 
The difference of the ratio between the extracted and the input values is consistent with zero and the 
width of the pull distributions is one. 

In order to determine the ratio of branching fractions, the ratio of the yields of each decay mode must 
be corrected with the relative reconstruction efficiency. 
The relative reconstruction efficiency depends in turn on the different decay-in-flights and nuclear 
interaction probabilities of the kaon and pion from the two decay modes and on the slightly different 
track momentum spectra. 
The relative reconstruction efficiency for the two decay modes is 
$\epsilon_{rel}=\mbox{0.991}\pm\mbox{0.005}$ which is derived from the Monte Carlo simulation. 
 
In this analysis, we use a Monte Carlo simulation to parameterize the $\alpha$ distributions of each 
signal and to determine the relative reconstruction efficiency for the two decay modes. 
The Monte Carlo generation proceeds as follows. 
Transverse momentum and rapidity distributions of single $b$ quarks are generated based on 
next-to-leading order (NLO) pertubative QCD calculation~\cite{nason}. 
$B$ meson kinematic distributions are obtained by simulating Peterson fragmentation~\cite{peterson} on 
quark-level distributions. 
Additional fragmentation particles, correlated $b\bar{b}$ production and the underlying event structure 
are not generated. 
The $B$ meson spectrum used in the Monte Carlo simulation is from the inclusive 
$B\rightarrow \jpsi X$ measurement~\cite{acosta}. 
The \textsc{cleomc} program~\cite{qqcleo} is used to decay $\Bp$ mesons into the final states of interest.
The simulation of the CDF II detector and trigger is based on a \textsc{geant}~\cite{brun} description. 

Since both decay modes of interest have almost identical decay topology and kinematics, most systematic 
uncertainties cancel in this ratio measurement, including uncertainties in total integrated 
luminosity and trigger and reconstruction efficiencies. 
Remaining systematic uncertainties come from the uncertainties in the shapes of the mass distribution, 
the parameterized PDFs in the $\alpha$ variable, and from the determination of the relative 
reconstruction efficiency. 
The largest systematic uncertainty originates from the unknown shape of the combinatorial background 
in the mass distribution. 
To estimate this effect, a second order polynomial function is considered as an alternative model for 
the shape of the background mass distribution. 
The modelling of the width of the invariant mass distribution is determined from momentum scale resolution
studies~\cite{korn}.
An alternative model from a simple Gaussian is to include an additional Gaussian for 
potential different momentum resolutions of tracks reconstructed in different detector geometry coverage. 
We replace a Gaussian with a double Gaussian for modeling each signal mass distribution and fit again 
to evaluate the uncertainty coming from the non-Gaussian tails in the $\Bpjk$ mass distribution.
The uncertainties in the function parameters describing the $\alpha$ PDFs , in 
Eq.~(\ref{eq:alphas}) and (\ref{eq:alphab}), generate an uncertainty for the ratio measurement. 
The contribution of this uncertainty is estimated by performing the fit by varying the 
parameterzation variables of the PDFs by the $\pm 1\sigma$, obtained from Monte Carlo simulation. 
The uncertainty in $\epsilon_{rel}$ originates from the uncertainties of the nuclear interaction and 
the material description in the detector simulation.
The \textsc{geant} simulation calculates nuclear interaction probabilities of $\approx$ 4\% for $\pi^{+}$, 
$\pi^{-}$ and $K^{-}$, and $\approx$ 3\% for $K^{+}$. 
We then assign a 25\% uncertainty to the calculated nuclear interaction probabilities as the 
uncertainty of the detector material description in the detector simulation, and take the resulting 
uncertainty in $\epsilon_{rel}$ as a systematic uncertainty. 
We determine the total systematic uncertainty of 3.0\% on the measurement by adding the individual 
uncertainties in quadrature, and the contributions from each source are summarized in 
Table~\ref{tab:table1}.

\begin{table}[htbp]
\caption{\label{tab:table1}Summary of systematic uncertainties for
the ratio of branching fractions, $\ratio$.}
\begin{ruledtabular}
\begin{tabular}{cc}
Source                              &Uncertainty of \\
                                    &the ratio (\%) \\
\hline
Background Shape                    &   2.5         \\
Non-Gaussian tail of $\Bpjk$	    &   1.2         \\
$\alpha$ PDFs Parametrization       &   1.0         \\
Relative Reconstruction Efficiency  &   0.5         \\
\hline
Total Uncertainty                   &   3.0         \\
\end{tabular}
\end{ruledtabular}
\end{table}

From Eq.~(\ref{eq:ratio}), we derived the ratio of branching fractions, 
$$
\frac{\mathcal{B}(\Bpjp)}{\mathcal{B}(\Bpjk)} =
(\mbox{4.86} \pm \mbox{0.82(stat.)} \pm \mbox{0.15(syst.)})\%,
$$
where the first error is statistical and the second is systematic.

In conclusion, we present the measurement of the ratio of branching fractions between $\Bpjp$ and 
$\Bpjk$. 
This result is consistent with theoretical expectations and the previous measurements, and will improve 
the present world average (5.3$\pm$0.4)\%~\cite{pdg}.

We thank the Fermilab staff and the technical staffs of the participating institutions
for their vital contributions. This work was supported by the U.S. Department of Energy
and National Science Foundation; the Italian Istituto Nazionale di Fisica Nucleare; the
Ministry of Education, Culture, Sports, Science and Technology of Japan; the Natural
Sciences and Engineering Research Council of Canada; the National Science Council of the
Republic of China; the Swiss National Science Foundation; the A.P. Sloan Foundation; the
Bundesministerium f\"ur Bildung und Forschung, Germany; the Korean Science and Engineering
Foundation and the Korean Research Foundation; the Particle Physics and Astronomy Research
Council and the Royal Society, UK; the Russian Foundation for Basic Research; the
Comisi\'on Interministerial de Ciencia y Tecnolog\'{\i}a, Spain; in part by the European
Community's Human Potential Programme under contract HPRN-CT-2002-00292; and the Academy
of Finland.


\begin{thebibliography}{99}
\bibitem{geom}
    CDF II uses a cylindrical coordinate system in which $\phi$ is the azimuthal angle,
$r$ is the radius from the nominal beam line, and $z$ points in the beam direction, with
the origin at the center of the detector. The $r-\phi$ plane is the transverse plane perpendicular 
to the z axis.
\bibitem{Mass-alpha}
	$mathcal{M}_{B}^{\prime 2}(\alpha)   =  M_{B}^{2}+(M_{K}^{2}-M_{\pi}^{2})+
	2 E_{J/\psi}(\sqrt{M_K^2+P_K^2} - \sqrt{M_{\pi}^2+P_K^2}) 
	\simeq M_{B}^{2}+(1+\alpha)(M_{K}^{2}-M_{\pi}^{2})$, 
	assuming  $p_K>>M_K$ and  $p_K>>M_{\pi}$. All momentums are cacluated in Lab frame.
\end{thebibliography}
\end{document}